\documentclass[prl,onecolumn,amsmath,amssymb,superscriptaddress]{revtex4-1}
\usepackage{bm,psfrag,graphicx,setspace}
\usepackage[T1]{fontenc}
\usepackage[left]{lineno}
\selectfont

\begin{document}
\title{Active tuning of hybridized modes in a heterogeneous photonic molecule}
\author{Kevin C. Smith}
\thanks{These authors contributed equally to this Letter}
\affiliation{Department of Physics, University of Washington, Seattle, Washington 98195, USA}
\author{Yueyang Chen}
\thanks{These authors contributed equally to this Letter}
\affiliation{Department of Electrical and Computer Engineering, University of Washington, Seattle, Washington 98195, USA}
\author{Arka Majumdar}
\email{arka@uw.edu}
\affiliation{Department of Electrical and Computer Engineering, University of Washington, Seattle, Washington 98195, USA}
\affiliation{Department of Physics, University of Washington, Seattle, Washington 98195, USA}
\author{David J. Masiello}
\email{masiello@uw.edu}
\affiliation{Department of Chemistry, University of Washington, Seattle, Washington 98195, USA}

\begin{abstract}
From fundamental discovery to practical application, advances in the optical and quantum sciences rely upon precise control of light-matter interactions. Systems of coupled optical cavities are ubiquitous in these efforts, yet design and active modification of the hybridized mode properties remains challenging. In this Letter, we demonstrate the ability to thermally control the degree of hybridization in a heterogeneous photonic molecule composed of a ring resonator strongly coupled to a nanobeam photonic crystal cavity. Combining theory and experiment, we show that the composition of the resulting super-modes can be actively tailored and we derive temperature-dependent analytic expressions for the super-mode profiles, frequencies, and volumes. This work illustrates the potential for actively tunable, designer photonic properties using heterogeneous optical cavity devices.
\end{abstract}
\maketitle

Coupled optical microcavities serve as a basic building block for many integrated photonic systems and technologies. Similar to the way bound electronic states of individual atoms couple to form those of a molecule, confined photonic excitations of two or more optical cavities can electromagnetically interact to form so-called ``photonic molecules'' \cite{Bayer1998, Mukaiyama1999, Atlasov2008, Atlasov2011, Majumdar2012a, Zhang2018}. Electronic excitations in molecules are described through hybridization of the orbitals of the constituent atoms and, in analogy, the electromagnetic super-modes of photonic molecules can be constructed by blending the resonances of the individual cavities. While single cavities are instrumental to a diverse set of applications ranging from single photon generation \cite{Santori2002, Kuhn2002, Chang2006, Birnbaum2005} and strong light-matter coupling \cite{Weisbuch1992, Lidzey1998, Reithmaier2004, Peter2005} to sensing \cite{McKeever2004, Fischer2002, Oezdemir2014, Vollmer2008, Vollmer2012, Zhu2010, Vollmer2008a, He2011, Heylman2016} and cavity-controlled chemistry \cite{Herrera2016, Dunkelberger2016, Thomas2019, Schaefer2019, Du2019, Lather2019}, systems of two or more cavities have shown promise in a number of applications, including low-threshold lasing \cite{Nakagawa2005, Boriskina2006a, Smotrova2006}, cavity optomechanics \cite{Jiang2009, Hu2013, Cao2016}, nonclassical light generation \cite{Liew2010, Bamba2011, Gerace2014, Flayac2017, Dousse2010, Saxena2019, Choi2019}, quantum simulation \cite{Angelakis2007, Underwood2012, Majumdar2012, Georgescu2014, Hartmann2016}, and biochemical sensing \cite{Boriskina2006, Boriskina2010}.


Critical to the advantages of photonic molecules over individual cavities is the ability to engineer designer super-modes with properties that differ from those of the constituent components. Of particular interest are coupled cavity structures whose optical properties evolve with tunable parameters such as cavity-cavity separation and detuning. In recent years, the active tuning of such photonic molecules has been demonstrated in several experiments \cite{Zhang2018, Peng2014, Cao2016}, but all have focused on coupled structures composed of near-identical individual cavities. While these devices are useful for many applications, homogeneity of the constituent cavities limits the dynamic range of the resulting super-mode properties such as the mode volume, important both for the scaling of light-matter coupling and Purcell enhancement.

In contrast, a heterogeneous photonic molecule composed of two distinctly different cavities allows for a richer set of emergent properties with a wider scope of applications, such as improved single photon indistinguishability of quantum emitters \cite{Choi2019, Saxena2019}. However, lack of a theoretical framework analogous to molecular orbital theory that is capable of elucidating the dependencies of the composite system upon single cavity parameters makes design and analysis of coupled optical cavities difficult. Absent such a formalism, prediction of super-mode field profiles and other downstream properties such as hybridized resonant frequencies and mode volumes must be left to numerical simulation.  The latter can be costly for all but the simplest coupled cavities and impossible for many heterogeneous systems, providing impetus for theoretical advances in understanding cavity mode hybridization.

In this Letter, we demonstrate thermally tunable hybridization of optical cavity modes in a heterogeneous photonic molecule composed of a ring resonator and a nanobeam photonic crystal (PhC) cavity. This is achieved by embedding the coupled cavity structure in a high thermo-optic coefficient polymer that preferentially blue-shifts the nanobeam resonance relative to the ring due to the ``air-mode'' design of the PhC cavity \cite{Chen2019}. To better understand the resulting super-modes of this heterogeneous optical system, we introduce a theoretical framework which provides rigorous underpinnings to the more familiar coupled mode theory for hybridized cavity systems and, for the first time, derive analytic expressions for the super-mode field profiles and mode volumes expressed in terms of the single cavity field profiles. Using this formalism, we demonstrate the ability to extract crucial system parameters, such as the bare resonant frequencies and couplings, as a function of the temperature-dependent detuning. Lastly, we use this theory to predict the evolution of the resonant frequencies, field profiles, and hybridized mode volumes of the two super-modes, revealing a temperature-dependent progression which spans a full order of magnitude and results in the coalescence of the two mode volumes near zero detuning.

\begin{figure}
\includegraphics[width=0.95\textwidth]{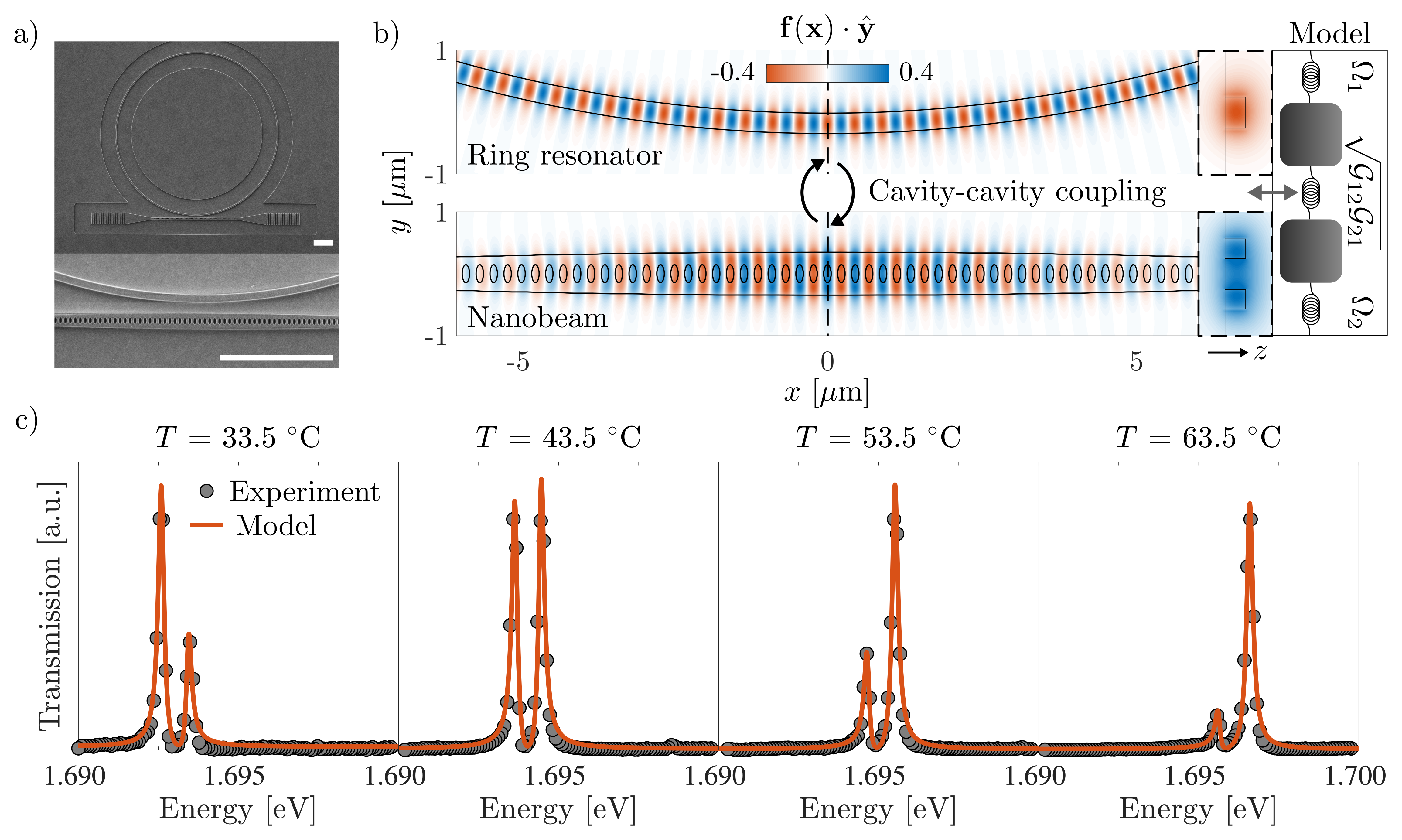}
\caption{\label{f1} (a) SEM image of the SU-8 cladded, coupled ring resonator-nanobeam device with a 500 nm gap between ring and nanobeam at the point of closest separation. Scale bar: $5$ $\mu$m. (b) $y$-component of the electric field profiles for the nanobeam cavity mode (bottom) and ring resonator mode (top) studied. The system is modeled as a coupled oscillator, parameterized by an effective coupling strength $\sqrt{\mathcal{G}_{12}\mathcal{G}_{21}}$ and effective frequencies $\Omega_i$ distinct from the bare resonant frequencies $\omega_i$. (c) Transmission spectra collected for four equally-spaced temperatures (gray circles) with simultaneous least-squares fits to the model overlaid (red lines).}
\end{figure}

Fig. \ref{f1}a. displays a scanning electron microscope (SEM) image of the heterogeneous, coupled cavity system fabricated on a 220 nm thick silicon nitride film, grown on thermal oxide on a silicon substrate. The pattern is defined by e-beam lithography and reactive ion etching \cite{Fryett2018}. The nanobeam cavity is designed such that a significant portion of the cavity field is concentrated in SU-8 polymer, which both forms a cladding for the entire device and fills the holes of the PhC \cite{Chen2019} (see Fig. \ref{f1}b). In contrast, the ring resonator mode is predominantly confined within the silicon nitride. Due to the relatively high thermo-optic coefficient of the polymer ($\sim -10^{-4}/^{\circ}\textrm{C}$), which is nearly an order of magnitude larger than that of silicon nitride, heating the entire device leads to a blue-shift of the nanobeam cavity mode relative to that of the ring. The detuning between the ring and nanobeam modes can therefore be reversibly controlled by changing the temperature.

To investigate the effect of ring-nanobeam mode detuning, the transmission spectrum is measured through the nanobeam PhC cavity for a range of temperatures spanning $33.5-73.5^\circ\textrm{C}$. Spectra are measured using a supercontinuum laser which is coupled to the system via an on-chip grating (see Fig. 1a). The transmitted light is collected through the opposite grating and is sent to the spectrometer. While the gratings already provide a spatial separation to improve the signal to noise ratio, a pinhole is used in the confocal microscopy setup to collect light only from the output grating. The temperature of the entire chip is controlled using a hot-plate. Fig. \ref{f1}c displays the resulting transmission spectra (gray circles) for a subset of temperatures, with additional measurements included in the Supplemental Material \cite{SM}. As the cavity modes of the ring and nanobeam are coupled, it is difficult to distinguish how much of the energy separation between transmission peaks at each temperature is due to detuning versus mode splitting resulting from coupling.


Understanding the impact of these individual contributions and analysis of emergent properties requires a theoretical formalism capable of describing the super-modes of the coupled ring-nanobeam structure. Coupled mode theory provides one such approach, but typically relies on several phenomenological rates which simplify modeling, often at the expense of over-simplifying the underlying physics. Furthermore, coupled mode theory does not provide a means to predict super-mode properties of interest for heterogeneous photonic molecules, such as hybridized mode volumes. To amend these deficiencies we develop a first principles theory that provides analytic understanding of the super-mode resonant frequencies, field profiles, and volumes based only upon knowledge of the individual, uncoupled cavities.

The resonant modes of an optical cavity are given by the independent harmonic solutions of the wave equation
\begin{equation}
\nabla\times\nabla\times\mathbf{A}(\mathbf{x},t) + \frac{\varepsilon(\mathbf{x})}{c^2}\ddot{\mathbf{A}}(\mathbf{x},t)=\mathbf{0},
\label{eq:waveeq}
\end{equation}
where $\mathbf{A}$ is the vector potential related to the cavity fields by the usual relations $\mathbf{E} = -\dot{\mathbf{A}}/c$ and  $\mathbf{B} = \nabla\times\mathbf{A}$, $\varepsilon(\mathbf{x})$ is the dielectric function of the structure of interest, and $c$ is the speed of light. As is typical for cavity quantum electrodynamics calculations, we work entirely in the generalized Coulomb gauge defined by $\nabla\cdot\varepsilon(\mathbf{x})\mathbf{A}(\mathbf{x})=0$ which leads to a vanishing scalar potential for systems without free charge \cite{Glauber1991, Dalton1996}. While optical cavities may alternatively be described at the level of the fields themselves, the vector potential accommodates a more natural basis for both a Lagrangian formulation of the cavity dynamics and canonical quantization \cite{cohen1997photons}.

Given $\varepsilon(\mathbf{x})$, it is in principle straightforward to numerically solve for the modes of the two-cavity structure in Fig. \ref{f1}a. Such an approach, however, offers limited predictivity and insight into the interaction between the individual ring resonator and nanobeam modes. In addition, the vastly different length scales of the ring resonator and nanobeam cavity make electromagnetic simulations of the coupled structures computationally challenging, rendering a purely numerical exploration of parameter space infeasible. A more flexible strategy is to numerically solve for the modes of the individual, uncoupled cavities. With the aid of analytics, these individual modes may then be appropriately mixed to form super-modes dependent on basic system parameters such as the spectral detuning and physical separation between the cavities.

Considering just a single cavity mode of both the ring resonator and nanobeam, the vector potential for the double cavity structure can be expanded as
\begin{equation}
\mathbf{A}(\mathbf{x},t) = \sum_{i=1,2}\frac{\sqrt{4\pi}c}{V_i}q_i(t)\mathbf{f}_i(\mathbf{x}).
\label{eq:expansion}
\end{equation}
Here, $i=1,2$ corresponds to the ring and nanobeam, respectively, while $\mathbf{f}_i(\mathbf{x})$ is a mode function of the $i$th cavity \cite{SM} and $q_i(t)$ a time-dependent amplitude. The former are normalized such that the mode volume is given by $V_i = \int d^3x \, \varepsilon_i(\mathbf{x})\left|\mathbf{E}_i(\mathbf{x})\right|^2/\textrm{max}[\varepsilon_i(\mathbf{x})\left|\mathbf{E}_i(\mathbf{x})\right|^2]=\int d^3x \, \varepsilon_i(\mathbf{x})\left|\mathbf{f}_i(\mathbf{x})\right|^2.$
The mode expansion in Eq. (\ref{eq:expansion}) is approximate and, in general, requires additional terms to ensure Gauss's law is obeyed \cite{Johnson2002, cavitycavity_unpub}. However, these contributions only become physically relevant at small inter-cavity separations where the evanescent field of one cavity ``spills'' into the dielectric medium composing the other, and therefore may be ignored for the ring-nanobeam resonator studied \cite{SM}.


The resonant super-mode frequencies are most easily computed through diagonalization of the equations of motion for the generalized coordinates $q_i$. Deriving such equations is straightforward using standard techniques of Lagrangian mechanics \cite{SM}, but an equivalent route involves directly integrating Eq. (\ref{eq:waveeq}) \cite{Yariv1999}. Regardless of the approach, the coupled equations of motion are



\begin{equation}
\begin{split}
\frac{d^2}{dt^2}\left[\begin{matrix}q_1 \\ q_2\end{matrix}\right] = \left[\begin{matrix}\Omega_1^2 & \mathcal{G}_{12} \\ \mathcal{G}_{21} & \Omega_2^2\end{matrix} \right]\left[\begin{matrix}q_1 \\ q_2\end{matrix}\right],
\label{eq:eom}
\end{split}
\end{equation}
where $\Omega_i^2 =(\bar{\omega}_i^2 - \bar{g}_E\bar{g}_M)/(1-\bar{g}_E^2/\bar{\omega}_1\bar{\omega}_2)$ and $\mathcal{G}_{ij} = \sqrt{\bar{\omega}_j\bar{V}_i/\bar{\omega}_i\bar{V}_j}\left(\bar{\omega}_i\bar{g}_M - \bar{\omega}_j\bar{g}_E\right)/(1-\bar{g}_E^2/\bar{\omega}_1\bar{\omega}_2)$ define effective resonant frequencies and couplings.

These coupled equations of motion differ from those often assumed in application of coupled mode theory to multiple cavity systems \cite{haus1984waves, Li2010, Zhang2018}. In particular, the diagonal elements of the above coefficient matrix are distinct from the bare resonance frequencies $\omega_i$. This is a consequence of the absence of a weak coupling approximation, resulting in coupling-induced resonance shifts \cite{Popovic2006} that scale as higher-order products of the three distinct coupling parameters corresponding to the electric ($g_E$) and magnetic ($g_M$) inter-cavity couplings, and the polarization-induced intra-cavity self-interaction ($\Sigma_i$). The former are defined by $g_E= \sqrt{{\omega_1\omega_2}/{V_1V_2}}\int d^3x \,\varepsilon(\mathbf{x})\mathbf{f}_1(\mathbf{x})\cdot\mathbf{f}_2(\mathbf{x})$ and $g_M = \sqrt{1/{\omega_1\omega_2V_1V_2}}\int d^3x \,[\omega_1^2\varepsilon_1(\mathbf{x})+\omega_2^2\varepsilon_2(\mathbf{x})]\mathbf{f}_1(\mathbf{x})\cdot\mathbf{f}_2(\mathbf{x}),$ while $\Sigma_i={\sqrt{1/V_1V_2}}\int d^3x \,[\varepsilon(\mathbf{x}) - \varepsilon_i(\mathbf{x})]\left|\mathbf{f}_i(\mathbf{x})\right|^2$ does not explicitly appear in Eq. (\ref{eq:eom}). Instead, all inter-cavity couplings, resonant frequencies ($\omega_i$), and mode volumes ($V_i$) have been replaced by renormalized counterparts (indicated by a bar), defined explicitly in the Supplemental Material \cite{SM}.



While coupled mode theory often reduces cavity-mode interactions to a single coupling parameter independent of the detuning, we note that this is not completely accurate, and more rigorous first-principles treatments relying on tight-binding methods \cite{Yariv1999, Bayindir2000} have revealed three distinct coupling parameters in agreement with those defined above. However, as shown in Eq. (\ref{eq:eom}), these three parameters may be combined, along with the resonant frequencies, to form effective coupled oscillator equations which account for these subtleties. Notably, all parameters may be computed given only the dielectric function composing the individual cavities along with associated field mode profiles.

Aided by the effective oscillator equations in Eq. (\ref{eq:eom}), the transmission spectrum is derived through standard input-output methods \cite{haus1984waves, Collett1984, SM}, yielding
\begin{equation}
\mathcal{T}(\omega) = \left|\frac{\kappa}{\omega-\Omega_1 + i\kappa+\frac{\mathcal{G}_{12}\mathcal{G}_{21}/4\Omega_1\Omega_2}{\omega-\Omega_2}}\right|^2.
\label{eq:transmission}
\end{equation}
Simultaneous least-squares fits are performed to transmission spectra at the eight experimentally probed temperatures shown in Fig. \ref{f1}c and the Supplemental Material \cite{SM}. To minimize the number of free parameters, $\Sigma_1$, $\Sigma_2$, $V_1$ and $V_2$ are calculated using the theory, supplemented by numerically calculated single cavity field profiles. Similarly, $g_E$ and $g_M$ are constrained to within $\pm 1 \%$ of their theoretical values, while the waveguide-induced dissipation rate $\kappa$ is estimated from electromagnetic simulation of the nanobeam.

The remaining free parameters, displayed in the top row of Table \ref{t1}, are extracted through a simultaneous least-squares fit to all measured transmission spectra. Among them is the resonant frequency of both the ring resonator and nanobeam at room temperature $T_0$ and associated intrinsic dissipation rates, the latter of which may be introduced via input-output theory in the standard way by generalizing $\Omega_1$ and $\Omega_2$ to be complex-valued \cite{haus1984waves}. We find that the temperature dependence of the resonant wavelength of each cavity is well-approximated as linear. All other parameters are assumed to depend negligibly upon temperature and are treated as constant. Even with these simplifying approximations, agreement between experiment (circles) and theory (solid lines) is excellent, as evident in Fig. \ref{f1}c.

\begin{table}
\caption{Parameter Estimates} 
\centering 
\setlength{\tabcolsep}{0.5em}
\def\arraystretch{2}
\begin{tabular}{c c c c c c c} 
\hline\hline 
$\hbar\omega_1(T_0)$ & $\hbar\omega_2(T_0)$ & $d\lambda_1/dT$ & $d\lambda_2/dT$ & $\hbar\gamma_1$ & $\hbar\gamma_2$ \\ [0.5ex] 
\hline 
1.6922 eV & 1.6918 eV & $-39$ pm/$^\circ$nC & $-50$ pm/$^\circ$nC & 0.16 meV & 0.23 meV  \\[0.5ex] 
\hline\hline 
$V_1$ & $V_2$ & $\kappa$ & $\hbar g_E$ & $\hbar g_M$ & $\Sigma_1$ & $\Sigma_2$  \\ [0.5ex]
\hline
5.0 $\mu$m$^3$ & 0.49 $\mu$m$^3$ & 9.7 $\mu$eV & $-16.4$ meV & $-15.6$ meV & 1.1$\times10^{-5}$ & 8.5$\times10^{-5}$  \\ [0.5ex]
\hline
\end{tabular}
\label{t1}
\end{table}

\begin{figure}
\includegraphics[width=0.8\textwidth]{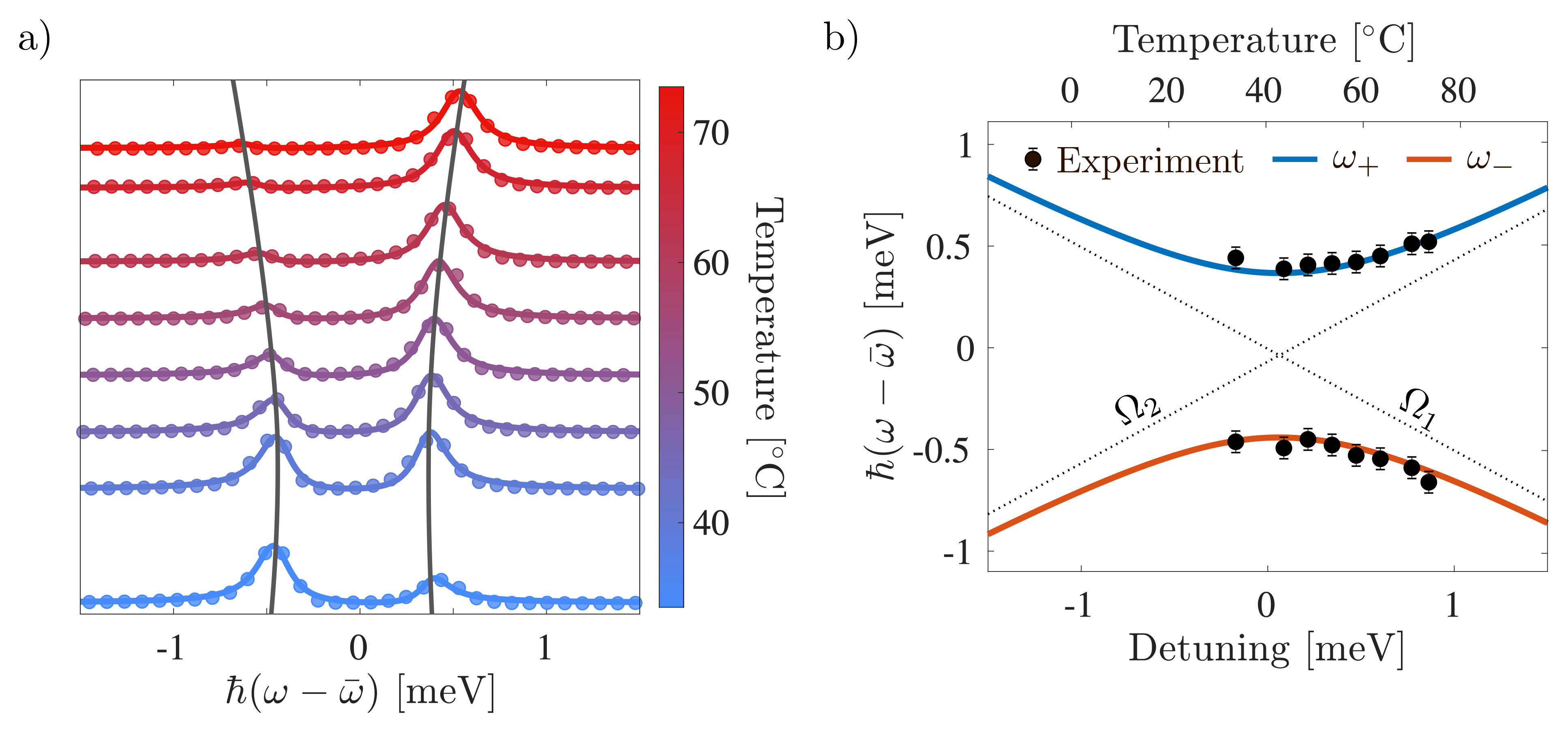}
\caption{\label{f2} (a) Anticrossing resulting from strong coupling between the ring resonator and nanobeam cavity modes. Experimental data are shown as circles, while colored solid lines display the resulting least-squares fit to Eq. (\ref{eq:transmission}). Gray lines overlay the theoretical values of $\omega_{\pm}$, extrapolated via parameter values obtained from the fits. (b) Evolution of the super-mode resonant frequencies as a function of detuning. Black points correspond to experimentally measured peak transmission energies, while error bars indicate uncertainty in the peak energy due to the finite density of transmission energies measured. Solid curves display theoretical super-mode energies computed from Eq. (\ref{eq:eom}), parameterized through simultaneous fits to transmission measurements.}
\end{figure}

Fig. \ref{f2}a displays the full set of transmission measurements (circles) and fits (curves) for all eight probed temperatures, while Fig. \ref{f2}b shows the super-mode resonant frequencies ($\omega_\pm$) as a function of energy detuning $\hbar\omega_2-\hbar\omega_1$. For each temperature measured, resonant frequencies are estimated from the peaks in transmission spectra and are shown as black circles. Theory curves (red and blue) are computed through diagonalization of the effective oscillator model in Eq. (\ref{eq:eom}) which we parameterize according to Table \ref{t1}. Because both ring and nanobeam modes blue-shift with increasing temperature, plotted curves and points are shifted with respect to the average resonant energy $\bar{\omega}=(\omega_+ +\omega_-)/2$ for both panels.

The resonant frequencies undergo an anticrossing as the system nears zero detuning around $T=40$ $^{\circ}$C, with upper and lower cavity polariton energies differing by $\sim0.8$ meV. Because the coupled oscillator model is parameterized by the effective frequencies $\Omega_1$ and $\Omega_2$, and not the bare cavity resonances $\omega_1$ and $\omega_2$, the anticrossing occurs where the former, and not the latter, are co-resonant. Thus, the anticrossing in Fig. \ref{f2} is slightly shifted from zero detuning. In addition, the super-mode resonances $\omega_{\pm}$ tend towards the effective frequencies (dotted lines) at large positive and negative values of the detuning. Strong coupling is confirmed quantitatively through comparison of the computed effective coupling strength \cite{SM} with the dissipation rates reported in Table \ref{t1} \cite{Rodriguez2016, Novotny2010}. In particular, we find that $\left|\hbar\sqrt{\mathcal{G}_{12}\mathcal{G}_{21}/4\Omega_1\Omega_2}\right|\approx0.40$ meV, nearly double the dominant intrinsic dissipation rate $\hbar\gamma_1=0.23$ meV.

Hybridization is further investigated through inspection of the super-mode profiles
\begin{equation}
\begin{split}
\mathbf{f}_{\mp}(\mathbf{x}) &= \frac{1}{A(\theta)}\left[\left(\frac{\mathcal{G}_{12}}{\mathcal{G}_{21}}\right)^{1/4}\sqrt{\frac{V_2}{V_1}}\mathbf{f}_1(\mathbf{x})\cos\theta  - \left(\frac{\mathcal{G}_{21}}{\mathcal{G}_{12}}\right)^{1/4}\sqrt{\frac{V_1}{V_2}}\mathbf{f}_2(\mathbf{x})\sin\theta\right]  \\
\mathbf{f}_{\pm}(\mathbf{x}) &= \frac{1}{B(\theta)}\left[\left(\frac{\mathcal{G}_{21}}{\mathcal{G}_{12}}\right)^{1/4}\sqrt{\frac{V_1}{V_2}}\mathbf{f}_2(\mathbf{x})\cos\theta +\left(\frac{\mathcal{G}_{12}}{\mathcal{G}_{21}}\right)^{1/4}\sqrt{\frac{V_2}{V_1}}\mathbf{f}_1(\mathbf{x})\sin\theta\right]  \\
\end{split}
\label{eq:modefunc}
\end{equation}
and their associated mode volumes
\begin{equation}
\begin{split}
V_\mp&=V_1\left[\frac{V_2}{V_1}\sqrt{\frac{\mathcal{G}_{12}}{\mathcal{G}_{21}}}\frac{1+\Sigma_1}{A(\theta)^2}\right]\cos^2\theta + V_2\left[\frac{V_1}{V_2}\sqrt{\frac{\mathcal{G}_{21}}{\mathcal{G}_{12}}}\frac{1+\Sigma_2}{A(\theta)^2}\right]\sin^2\theta - \sqrt{V_1V_2}\left[\frac{g_E/\sqrt{\omega_1\omega_2}}{A(\theta)^2}\right]\sin2\theta  \\
V_\pm&=V_2\left[\frac{V_1}{V_2}\sqrt{\frac{\mathcal{G}_{21}}{\mathcal{G}_{12}}}\frac{1+\Sigma_2}{B(\theta)^2}\right]\cos^2\theta + V_1\left[\frac{V_2}{V_1}\sqrt{\frac{\mathcal{G}_{12}}{\mathcal{G}_{21}}}\frac{1+\Sigma_1}{B(\theta)^2}\right]\sin^2\theta + \sqrt{V_1V_2}\left[\frac{g_E/\sqrt{\omega_1\omega_2}}{B(\theta)^2}\right]\sin2\theta,
\end{split}
\label{eq:modevol}
\end{equation}
where $A(\theta)$ and $B(\theta)$ are normalization factors \cite{SM}, $\theta=(1/2)\tan^{-1}(2\sqrt{\mathcal{G}_{12}\mathcal{G}_{21}}/[\Omega_2^2-\Omega_1^2])$ is the mixing angle, and the upper (lower) subscript corresponds to the case $\theta > 0$ ($\theta < 0$). The mixing angle has two distinct regimes; when the detuning is much larger than the effective coupling strength ($\theta\to0$), the above mode functions reduce to those of the bare ring resonator and nanobeam cavity. In contrast, for small detuning relative to the coupling ($\theta\to\pm\pi/4$) the mode functions become a superposition of $\mathbf{f}_1(\mathbf{x})$ and $\mathbf{f}_2(\mathbf{x})$.

\begin{figure}
\includegraphics[width=0.95\textwidth]{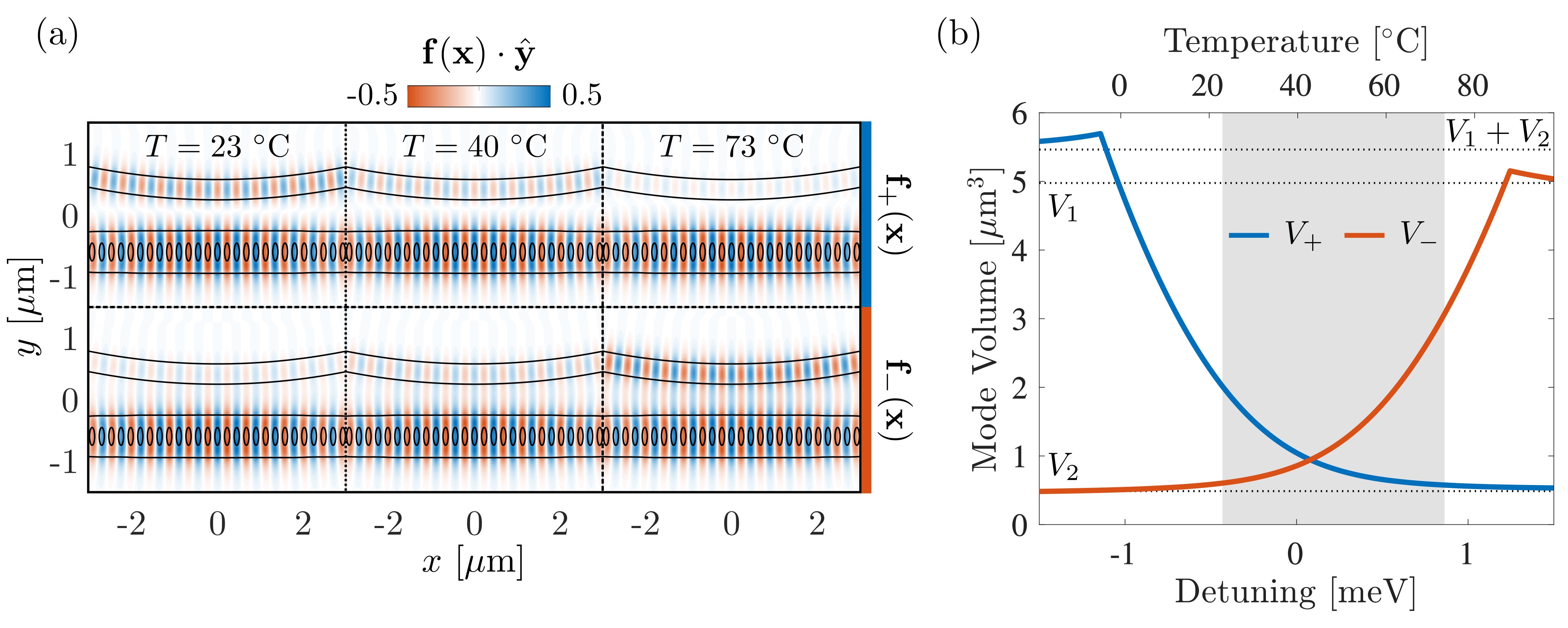}
\caption{\label{f3}(a) Field profile for the upper (top) and lower (bottom) cavity polaritons at various temperatures. Both super-modes are dominated by the nanobeam field at all observed temperatures due to the weighting of $\mathbf{f}_1(\mathbf{x})$ and $\mathbf{f}_2(\mathbf{x})$ in Eq. (\ref{eq:modefunc}). (b) Hybridized mode volumes $V_+$ (blue curve) and $V_-$ (red curve) of the upper and lower cavity polaritons. The gray region indicates the range of experimentally measured temperatures, while dotted lines specify $V_1$, $V_2,$ and $V_1+V_2$. Due to the predominant localization of both modes in the nanobeam cavity, both $V_+$ and $V_-$ coalesce at a value less than 5 times the mode volume of the isolated ring resonator mode.}
\end{figure}

Fig. \ref{f3}a shows the evolution of the $y$-component of the upper (top) and lower (bottom) cavity polariton field profiles across the experimentally measured temperature range. Because the limits of this range constrain the mixing angle to $-\pi/8\lesssim \theta \lesssim \pi/6$, neither $\mathbf{f}_+(\mathbf{x})$ nor $\mathbf{f}_-(\mathbf{x})$ entirely localize to one of the constituent cavities at any probed temperature. For all mode profiles shown, a significant portion of the field is contributed by the mode function of the nanobeam $\mathbf{f}_2(\mathbf{x})$. We note, however, that there is no fundamental reason that the device could not be heated past the maximum temperature studied here ($73$ $^\circ$C), or cooled below room temperature.

Notably, the super-mode profiles are not equal superpositions of $\mathbf{f}_1(\mathbf{x})$ and $\mathbf{f}_2(\mathbf{x})$ near zero detuning ($T =40$ $^\circ$C). This may be understood by considering the large mismatch in mode volume between the ring resonator and nanobeam modes ($V_1/V_2 \sim 10 $). According to Eq. (\ref{eq:modefunc}), the nanobeam contribution to both $\mathbf{f}_+(\mathbf{x})$ and $\mathbf{f}_-(\mathbf{x})$ scales like $(V_1/V_2)^{1/4}$, while that of the ring resonator scales like  $(V_2/V_1)^{1/4}$. As a result, both super-modes are predominantly localized to the nanobeam.

Fig. \ref{f3}b shows theoretical predictions for the hybridized mode volumes as a function of temperature-controlled detuning, calculated using Eq. (\ref{eq:modevol}) paired with the experimentally-informed parameter values in Table \ref{t1}. As before, blue and red curves correspond to the upper and lower cavity polaritons in Fig. \ref{f2}a. The gray region indicates the range of experimentally probed temperatures. Both hybridized mode volumes tend towards those of the individual cavities at large positive and negative detuning and coalesce at a value of $V_\pm\approx 0.95$ $\mu\textrm{m}^3$, more than a factor of 5 less than the mode volume of the isolated ring resonator.

While the nanobeam mode volume $V_2$ clearly serves as a lower bound for $V_\pm$, analysis of Eq. (\ref{eq:modevol}) indicates a maximum near $V_1+V_2$. $V_+$ slightly exceeds this value due to constructive interference between the two modes, while $V_-$ peaks at a value below $V_1+V_2$ due to destructive interference. Both mode volumes display a ``turning point'' at values of the mixing angle $\theta$ such that $\mathbf{f}_1(\mathbf{x})$ and $\mathbf{f}_2(\mathbf{x})$ are equally-weighted in either $\mathbf{f}_+(\mathbf{x})$ or $\mathbf{f}_-(\mathbf{x})$. Due to the large mismatch between $V_1$ and $V_2$, between these two points is a full order-of-magnitude of attainable values for both hybridized mode volumes, illustrating the potential of this heterogeneous device for actively-tunable photonic properties.

In conclusion, we have demonstrated actively tunable hybridization in a heterogeneous photonic molecule consisting of a ring resonator coupled to a photonic crystal cavity. Aided by a theoretical formalism developed to study hybridized cavity states, we show the capability to extract system parameters from experiment, resulting in a predictive effective oscillator model distinct from those typically assumed by coupled mode theory. We leverage this model to derive analytic expressions for the super-mode resonant frequencies, field profiles, and mode volumes, and use it to elucidate their evolution with temperature. Finally, we show that thermally-tunable hybridized mode volumes spanning a full order of magnitude are possible in the coupled ring-nanobeam resonator system, highlighting the actively tunable, designer photonic properties offered by heterogeneous coupled cavities.


\begin{acknowledgments}
We thank Dr. Xiang-Tian Kong for helpful discussions pertaining to the electromagnetic simulations. This research was supported by the National Science Foundation under the following awards: QII-TAQS-1936100 (Y.C., A.M., K.C.S., D.J.M.), CHE-1836500 (Y.C., A.M.), CHE-1664684 (K.C.S., D.J.M.), and CHE-1836506 (K.C.S., D.J.M.). A.M. also acknowledges support from the Sloan Foundation. All fabrication processes were performed at the Washington Nanofabrication Facility, a National Nanotechnology Coordinated Infrastructure site at the University of Washington, which is supported in part by funds from the National Science Foundation (through awards NNCI-1542101, 1337840 and 0335765), the National Institutes of Health, the Molecular Engineering \& Sciences Institute, the Clean Energy Institute, the Washington Research Foundation, the M. J. Murdock Charitable Trust, Altatech, ClassOne Technology, GCE Market, Google, and SPTS.
\end{acknowledgments}

\bibliography{../../../JabRef/MasterBib}
\end{document}


\title{Supplemental Material: Active tuning of hybridized modes in a heterogeneous photonic molecule}
\author{Kevin C. Smith}
\thanks{These authors contributed equally to this Letter}
\affiliation{Department of Physics, University of Washington, Seattle, Washington 98195, USA}
\author{Yueyang Chen}
\thanks{These authors contributed equally to this Letter}
\affiliation{Department of Electrical and Computer Engineering, University of Washington, Seattle, Washington 98195, USA}
\author{Arka Majumdar}
\affiliation{Department of Electrical and Computer Engineering, University of Washington, Seattle, Washington 98195, USA}
\affiliation{Department of Physics, University of Washington, Seattle, Washington 98195, USA}
\author{David J. Masiello}
\affiliation{Department of Chemistry, University of Washington, Seattle, Washington 98195, USA}
\maketitle

\section{Full set of experimental spectra and fits}
\begin{figure}[h]
\includegraphics[width=0.95\textwidth]{SI_alltransmission.png}
\caption{\label{f1}}
\end{figure}

\section{Theoretical formalism}
\subsection{Single cavity}
\noindent As discussed in the main text, the wave equation for the vector potential
\begin{equation}
	\nabla\times\nabla\times\mathbf{A}(\mathbf{x},t) + \frac{\varepsilon(\mathbf{x})}{c^2}\ddot{\mathbf{A}}(\mathbf{x},t)=0.
\end{equation}
encodes the electromagnetic resonances of an arbitrary, dispersionless system described by the dielectric function $\varepsilon(\mathbf{x})$. Importantly, we work in the generalized Coulomb gauge \cite{Glauber1991, Dalton1996} defined by the condition $\nabla\cdot\mathbf{A}(\mathbf{x})\varepsilon(\mathbf{x})=0$. In analogy to the typical Coulomb gauge \cite{cohen1997photons}, this condition allows for the simplification $\phi(\mathbf{x})=0$ within the presence of bound matter described by $\varepsilon(\mathbf{x})$, but in the absence of free charge.

First considering a single isolated cavity, the vector potential may be expanded as
\begin{equation}
\mathbf{A}(\mathbf{x},t) = \sum_m\frac{\sqrt{4\pi}c}{V_m}q_m(t)\mathbf{f}_m(\mathbf{x})
\label{eq:singleexp}
\end{equation}
where $q_m(t)$ is a time-dependent amplitude, $\mathbf{f}_m(\mathbf{x})$ is the mode function, and $V_m$ the mode volume. The mode functions obey the following four properties:
\begin{enumerate}
\item {\emph{$\mathbf{f}_m(\mathbf{x})$ is a solution to the generalized Helmholtz equation:}}
\begin{equation}
	\nabla\times\nabla\times\mathbf{f}_m(\mathbf{x}) = \varepsilon(\mathbf{x})\frac{\omega_m^2}{c^2}\mathbf{f}_m(\mathbf{x})
\end{equation}
\item {\emph{Due to the modified Coulomb gauge condition, $\varepsilon(\mathbf{x})\mathbf{f}_m(\mathbf{x})$ is transverse:}} 
\begin{equation}
\nabla\cdot\varepsilon(\mathbf{x})\mathbf{f}_m(\mathbf{x})=0
\end{equation}
\item {\emph{$\mathbf{f}_m(\mathbf{x})$ is normalized such that} $\textrm{max} \{\varepsilon(\mathbf{x})\left|\mathbf{f}_m\right|^2\}=1$, \emph{and therefore naturally defines the mode volume}:}
\begin{equation}
V_m = \frac{\int d^3x\,\varepsilon(\mathbf{x})\left|\mathbf{E}_m\right|^2}{\textrm{max} \{\varepsilon(\mathbf{x})\left|\mathbf{E}_m\right|^2\}} = \int d^3x\,\varepsilon(\mathbf{x})\left|\mathbf{f}_m(\mathbf{x})\right|^2,
\end{equation}
\item {\emph{The set of mode functions $\{\mathbf{f}_m(\mathbf{x})\}$ form an orthogonal basis}:}
\begin{equation}
\int d^3x\, \varepsilon(\mathbf{x})\mathbf{f}_m(\mathbf{x})\cdot\mathbf{f}_n(\mathbf{x})=V_m\delta_{mn}
\end{equation}
\end{enumerate}

\subsection{Two-cavity photonic molecule}
While it is intuitive to expect that the total field in the two-cavity structure, it is easy to see that this is not generally true by appealing to Gauss's law. For example, imagine the coupling of two individual cavities. The vector potential associated with the field of first cavity may be expanded as
\begin{equation}
\mathbf{A}_1(\mathbf{x},t) = \sum_m\frac{\sqrt{4\pi}c}{V_{1m}}q_{1m}(t)\mathbf{f}_{1m}(\mathbf{x}),
\end{equation}
where $\nabla\cdot\varepsilon_1(\mathbf{x})\mathbf{f}_{1m}(\mathbf{x})=0$, per Gauss's law in the modified Coulomb gauge. Here, $\varepsilon_1(\mathbf{x})$ is the dielectric function describing the isolated single cavity. Analogously, the vector potential associated with the field of the second cavity may be expanded as 
\begin{equation}
\mathbf{A}_2(\mathbf{x},t) = \sum_m\frac{\sqrt{4\pi}c}{V_{2m}}q_{2m}(t)\mathbf{f}_{2m}(\mathbf{x}),
\end{equation}
where $\nabla\cdot\varepsilon_2(\mathbf{x})\mathbf{f}_{2m}(\mathbf{x})=0$ and $\varepsilon_2(\mathbf{x})$ is the dielectric function describing the second, isolated single cavity. One might think that vector potential associated with the field of the two-cavity photonic molecule, defined by $\varepsilon(\mathbf{x})$, may then be expanded as
\begin{equation}
	\mathbf{A}(\mathbf{x},t) = \sum_{i,m}\frac{\sqrt{4\pi}c}{V_{im}}q_{im}(t)\mathbf{f}_{im}(\mathbf{x}),
\end{equation}
where $i=1,2$. However, this cannot be generally true as $\nabla\cdot\varepsilon(\mathbf{x})\mathbf{A}(\mathbf{x},t)\neq 0$, and Gauss's law is therefore not obeyed. A simple fix is to modify the above mode functions, solved for in the single cavity case, by gauge transforming to the ``correct" basis such that $\nabla\cdot\varepsilon(\mathbf{x})\widetilde{\mathbf{f}}_{im}(\mathbf{x})=0$. Through appeal to Gauss's law, it may be shown that this is achieved by correcting the mode function as
\begin{equation}
\mathbf{f}_{im}(\mathbf{x})\rightarrow\widetilde{\mathbf{f}}_{im}(\mathbf{x}) = \mathbf{f}_{im}(\mathbf{x})+\nabla\psi_{im}(\mathbf{x}),
\end{equation}
where 
\begin{equation}
\nabla\cdot(\varepsilon(\mathbf{x})\nabla\psi_{im}(\mathbf{x}))=-\nabla\cdot(\varepsilon(\mathbf{x})-\varepsilon_i(\mathbf{x})).
\label{eq:genpoi}
\end{equation} 
It is clear from this expression that $\psi_{im}(\mathbf{x})$ is a solution to the generalized Poisson equation. On the right-hand side is a source term, corresponding to the polarization induced by the $m$th mode of the $i$th cavity, within the region that the dielectric function has changed from the single cavity to the two-cavity description. Taking into account these corrections, the vector potential may be expanded in the two-cavity case as
\begin{equation}
	\mathbf{A}(\mathbf{x},t) = \sum_{i,m}\frac{\sqrt{4\pi}c}{V_{im}}q_{im}(t)[\mathbf{f}_{im}(\mathbf{x})+\nabla\psi_{im}(\mathbf{x})].
\end{equation}

\subsection{$\nabla\psi_i(\mathbf{x})$ for the ring-nanobeam system}

Figs. S$\ref{f2}$ and S$\ref{f3}$ show a component-wise comparison between the unperturbed mode functions $\mathbf{f}_i(\mathbf{x})$ and the induced contribution $\nabla\psi_i(\mathbf{x})$, numerically solved using iterative techniques \cite{Fisicaro2016}. For both the nanobeam and the ring, the $\hat{\mathbf{x}}$ and $\hat{\mathbf{y}}$ components of $\nabla\psi_i(\mathbf{x})$ are a full two orders of magnitude smaller than those of the unperturbed mode functions $\mathbf{f}_i(\mathbf{x})$. Similarly, the $\hat{\mathbf{z}}$ component is a single order of magnitude smaller than the already nearly negligible $\hat{\mathbf{z}}$ component of $\mathbf{f}_i({\mathbf{x}})$. Therefore, these corrections are quite small for the $500$ nm cavity-cavity separation size considered, and have negligible effect on the parameter estimates presented in the main text. As a result, we take $\nabla\psi_i(\mathbf{x})$ to be vanishing as it dramatically simplifies calculations.

\begin{figure}
\centering
\includegraphics[width=0.8\textwidth]{SI_ring.png}
\caption{\label{f2} Induced corrections to the field profile of the ring resonator mode. The left column shows the field profile of the isolated ring, while the right shows the induced contributions $\nabla\psi_1(\mathbf{x})$ which are found by numerically solving Eq. (\ref{eq:genpoi}).}
\end{figure}

\begin{figure}
\centering
\includegraphics[width=0.8\textwidth]{SI_nanobeam.png}
\caption{\label{f3} Induced corrections to the field profile of the nanobeam mode. The left column shows the field profile of the isolated nanobeam, while the right shows the induced contributions $\nabla\psi_2(\mathbf{x})$.}
\end{figure}

\subsection{Dynamics for two single mode cavities}
\noindent Specializing to the case of the nanobeam mode interacting with a single ring resonator mode (and discarding the corrections $\nabla\psi_i$ as they are small for this system and will be expanded upon in future work \cite{cavitycavity_unpub}), the vector potential may be expanded as
\begin{equation}
	\mathbf{A}(\mathbf{x},t) = \sum_{i=1,2}\frac{\sqrt{4\pi}c}{V_{i}}q_{i}(t)\mathbf{f}_{i}(\mathbf{x}),
\end{equation}
As mentioned in the main text, equations of motions may be computed either through integration of the wave equation \cite{Yariv1999}, or via the standard Euler-Lagrange approach. Here, we follow the latter strategy and therefore use the standard electromagnetic Lagrangian
\begin{equation}
L= \int\frac{ d^3x}{8\pi}\,\left[\varepsilon(\mathbf{x})\frac{\dot{\mathbf{A}}^2}{c^2}+(\nabla\times\mathbf{A})^2\right]
\end{equation}
in the modified Coulomb gauge (and in the absence of free charge). Plugging in the above expansion for the vector potential leads to
\begin{equation}
\begin{split}
L = &\frac{1}{2}\sum_{i}\frac{\dot{q}_i}{V_i}\left[1+\Sigma_i\right]-\frac{1}{2}\sum_{i}\frac{\omega_i^2}{V_i}q_i^2 +\frac{g_E}{\sqrt{\omega_1\omega_2V_1V_2}}\dot{q}_1\dot{q}_2 - g_M\sqrt{\frac{\omega_1\omega_2}{V_1V_2}}q_1q_2,
\end{split}
\label{eq:L}
\end{equation}
where the analytic forms of $g_E$, $g_M$, and $\Sigma$ are given in the main text.
Application of the Euler-Lagrange equations then gives
\begin{equation}
\begin{split}
&\frac{\ddot{q}_1}{\bar{V}_1} + \bar{\omega}^2_1\frac{q_1}{\bar{V}_1} + \frac{\bar{g}_E}{\sqrt{\bar{\omega}_1\bar{\omega}_2\bar{V}_1\bar{V}_2}}\ddot{q}_{2}+\bar{g}_M\sqrt{\frac{\bar{\omega}_1\bar{\omega}_2}{\bar{V}_1\bar{V}_2}}q_2=0 \\
&\frac{\ddot{q}_2}{\bar{V}_2} + \bar{\omega}^2_2\frac{q_2}{\bar{V}_2} + \frac{\bar{g}_E}{\sqrt{\bar{\omega}_1\bar{\omega}_2\bar{V}_1\bar{V}_2}}\ddot{q}_{1}+\bar{g}_M\sqrt{\frac{\bar{\omega}_1\bar{\omega}_2}{\bar{V}_1\bar{V}_2}}q_1=0 \\
\end{split}
\end{equation}
where renormalized mode volumes, frequencies and coupling strengths are defined as
\begin{equation}
\begin{gathered}
\bar{V}_1 = V_1/(1+\Sigma_1) \htab \bar{V}_2 = V_2/(1+\Sigma_2) \\
\bar{\omega}_1 = \omega_1/\sqrt{1+\Sigma_1} \htab \bar{\omega}_2 = \omega_2/\sqrt{1+\Sigma_2}\\
\bar{g}_E = g_E/[(1+\Sigma_1)(1+\Sigma_2)]^{3/4} \\
\bar{g}_M = g_M/[(1+\Sigma_1)(1+\Sigma_2)]^{1/4}.
\end{gathered}
\end{equation}
Further algebra yields the equations of motion defined in the main text, 
\begin{equation}
\begin{split}
\frac{d^2}{dt^2}\left[\begin{matrix}q_1 \\ q_2\end{matrix}\right] = \left[\begin{matrix}\Omega_1^2 & \mathcal{G}_{12} \\ \mathcal{G}_{21} & \Omega_2^2\end{matrix} \right]\left[\begin{matrix}q_1 \\ q_2\end{matrix}\right]
\label{eq:eom}
\end{split}
\end{equation}
where $\Omega_i^2 =(\bar{\omega}_i^2 - \bar{g}_E\bar{g}_M)/(1-\bar{g}_E^2/\bar{\omega}_1\bar{\omega}_2)$ and $\mathcal{G}_{ij} = \sqrt{\bar{\omega}_j\bar{V}_i/\bar{\omega}_i\bar{V}_j}\left(\bar{\omega}_i\bar{g}_M - \bar{\omega}_j\bar{g}_E\right)/(1-\bar{g}_E^2/\bar{\omega}_1\bar{\omega}_2)$.

\subsection{Resonance energies, mode functions, and mode volumes}
Calculation of the normal mode resonance energies is achieved through diagonalization of the equations of motion in Eq. (\ref{eq:eom}). Denoting the coefficient matrix on the right-handside by $\mathbf{M}$, it may be written concisely in the form

\begin{equation}
\frac{d^2}{dt^2}\mathbf{q} =-\mathbf{M}\mathbf{q}
\label{eq:mateom}
\end{equation}
We next define the transformation matrix
\begin{equation}
\mathbf{X} = \mathbf{T} \mathbf{R}\mathbf{S}
\end{equation}
where
\begin{equation}
\begin{gathered}
\mathbf{T} = \left[\begin{matrix} (\mathcal{G}_{12}/\mathcal{G}_{21})^{1/4} & 0\\ 0 & (\mathcal{G}_{21}/\mathcal{G}_{12})^{1/4} \end{matrix} \right] , \htab
\mathbf{R} = \left[\begin{matrix} \cos{\theta} & -\sin{\theta}\\ \sin{\theta} & \cos{\theta} \end{matrix} \right],
\htab
\mathbf{S} = \left[\begin{matrix} \alpha_+ & 0\\ 0 & \alpha_- \end{matrix} \right]. \\
\end{gathered}
\end{equation}
Here, $\mathbf{T}$ and $\mathbf{R}$ together diagonalize $\mathbf{M}$, while $\mathbf{S}$ is an additional scaling which will be used to enforce a particular normalization condition for the mode functions.
Applying $\mathbf{X}^{-1}$ and $\mathbf{X}$ to the right to the left of Eq. \ref{eq:mateom}, we are left with
\begin{equation}
\begin{split}
\frac{d^2}{dt^2}\mathbf{X}^{-1}\mathbf{q} &=-\mathbf{X}^{-1}\mathbf{\Omega}\mathbf{X}\mathbf{X}^{-1}\mathbf{q} \\
\end{split}
\end{equation}
which results in the uncoupled equations of motion
\begin{equation}
\frac{d^2}{dt^2}\left[\begin{matrix} q_+\\ q_- \end{matrix}\right] =-\left[\begin{matrix}\omega_+^2 & 0 \\ 0 &  \omega_-^2\end{matrix}\right]\left[\begin{matrix} q_+\\ q_- \end{matrix}\right]
\end{equation}
where 
\begin{equation}
\begin{split}
\omega_+^2 =& \Omega_{1}\cos^2{\theta} + \Omega_{2}\sin^2{\theta} + 2\sqrt{\mathcal{G}_{12}\mathcal{G}_{21}}\sin{\theta}\cos{\theta}\\
\omega_-^2 =& \Omega_{2}\cos^2{\theta} + \Omega_{1}\sin^2{\theta} - 2\sqrt{\mathcal{G}_{12}\mathcal{G}_{21}}\sin{\theta}\cos{\theta}\\
\end{split}
\end{equation}
and the mixing angle $\theta$ is defined by
\begin{equation}
\theta = \frac{1}{2}\tan^{-1}\left(\frac{2\sqrt{\mathcal{G}_{12}\mathcal{G}_{21}}}{\Omega_{1}-\Omega_{2}}\right).
\end{equation}
Associated mode functions are most easily calculated by appealing to the vector potential expansion
\begin{equation}
\mathbf{A}(\mathbf{x},t) = \sqrt{4\pi}c\left[\frac{\mathbf{f}_1(\mathbf{x})}{V_1} \,\,\, \frac{\mathbf{f}_2(\mathbf{x})}{V_2}\right]\left[\begin{matrix}q_1(t) \\  q_2(t) \end{matrix}\right].
\end{equation}
Inserting the identity,
\begin{equation}
\mathbf{A}(\mathbf{x},t) = \sqrt{4\pi}c\left[\frac{\mathbf{f}_1(\mathbf{x})}{V_1} \,\,\, \frac{\mathbf{f}_2(\mathbf{x})}{V_2}\right]\mathbf{X}\mathbf{X}^{-1}\left[\begin{matrix}q_1(t) \\  q_2(t) \end{matrix}\right]
\end{equation}
and recognizing that 
\begin{equation}
\mathbf{X}^{-1}\left[\begin{matrix}q_1(t) \\  q_2(t) \end{matrix}\right] = \left[\begin{matrix}q_+(t) \\  q_-(t) \end{matrix}\right],
\end{equation}
it must then be true that
\begin{equation}
\left[\frac{\mathbf{f}_1(\mathbf{x})}{V_1} \,\,\, \frac{\mathbf{f}_2(\mathbf{x})}{V_2}\right]\mathbf{X} = \left[\frac{\mathbf{f}_+(\mathbf{x})}{V_+} \,\,\, \frac{\mathbf{f}_-(\mathbf{x})}{V_-}\right].
\label{eq:ftransform}
\end{equation}

Analogous to the single cavity case, it must be true that the following four properties are obeyed by the super-mode field profiles and their associated mode volumes:
\begin{enumerate}
\item {\emph{$\mathbf{f}_\pm(\mathbf{x})$ is a solution to the generalized Helmholtz equation:}}
\begin{equation}
	\nabla\times\nabla\times\mathbf{f}_\pm(\mathbf{x}) = \varepsilon(\mathbf{x})\frac{\omega_\pm^2}{c^2}\mathbf{f}_\pm(\mathbf{x})
\end{equation}
\item {\emph{$\varepsilon(\mathbf{x})\mathbf{f}_\pm(\mathbf{x})$ is transverse:}} 
\begin{equation}
\nabla\cdot\varepsilon(\mathbf{x})\mathbf{f}_\pm(\mathbf{x})=0
\end{equation}
\item {\emph{$\mathbf{f}_\pm(\mathbf{x})$ is normalized such that} $\textrm{max} \{\varepsilon(\mathbf{x})\left|\mathbf{f}_\pm\right|^2\}=1$, \emph{and therefore naturally defines the mode volume}:}
\begin{equation}
V_\pm = \frac{\int d^3x\,\varepsilon(\mathbf{x})\left|\mathbf{E}_\pm\right|^2}{\textrm{max} \{\varepsilon(\mathbf{x})\left|\mathbf{E}_\pm\right|^2\}} = \int d^3x\,\varepsilon(\mathbf{x})\left|\mathbf{f}_\pm(\mathbf{x})\right|^2,
\end{equation}
\item {\emph{The set of mode functions $\{\mathbf{f}_\pm(\mathbf{x})\}$ form an orthogonal basis}:}
\begin{equation}
\int d^3x \,\varepsilon(\mathbf{x})\mathbf{f}_\pm(\mathbf{x})\cdot\mathbf{f}_\mp(\mathbf{x})=0
\end{equation}
\end{enumerate}
where $\varepsilon(\mathbf{x})$ is now the full two-cavity dielectric function. The mode functions $\mathbf{f}_\pm(\mathbf{x})$ are unambiguously defined by Eq. (\ref{eq:ftransform}) up to an overall scaling factor, which is then set by choosing $\alpha_\pm$ such that Property 3 is obeyed. Carrying out the algebra leads to the resulting expressions for mode functions

\begin{equation}
\begin{split}
\mathbf{f}_+(\mathbf{x}) &= \frac{1}{A(\theta)}\left[  \left(\frac{\mathcal{G}_{12}}{\mathcal{G}_{21}}\right)^{1/4}\sqrt{\frac{V_2}{V_1}}\,\mathbf{f}_1(\mathbf{x})\cos\theta+ \left(\frac{\mathcal{G}_{21}}{\mathcal{G}_{12}}\right)^{1/4}\sqrt{\frac{V_1}{V_2}}\,\mathbf{f}_2(\mathbf{x})\sin\theta\right] \\ 
\mathbf{f}_-(\mathbf{x}) &= \frac{1}{B(\theta)}\left[  \left(\frac{\mathcal{G}_{21}}{\mathcal{G}_{12}}\right)^{1/4}\sqrt{\frac{V_1}{V_2}}\,\mathbf{f}_2(\mathbf{x})\cos\theta-\left(\frac{\mathcal{G}_{12}}{\mathcal{G}_{21}}\right)^{1/4}\sqrt{\frac{V_2}{V_1}}\,\mathbf{f}_1(\mathbf{x})\sin\theta\right] \\
\end{split}
\end{equation}
and their associated mode volumes
\begin{equation}
\begin{split}
V_+ &= V_1\left[\frac{V_2}{V_1}\left(\frac{\mathcal{G}_{12}}{\mathcal{G}_{21}}\right)^{\frac{1}{2}}\frac{1+\Sigma_1}{A(\theta)^2}\right]\cos^2\theta+ V_2\left[\frac{V_1}{V_2}\left(\frac{\mathcal{G}_{21}}{\mathcal{G}_{12}}\right)^{\frac{1}{2}}\frac{1+\Sigma_2}{A(\theta)^2}\right]\sin^2\theta + \sqrt{V_1V_2}\left[\frac{g_E/\sqrt{\omega_1\omega_2}}{A(\theta)^2}\right]\sin 2\theta \\
V_- &= V_1\left[\frac{V_2}{V_1}\left(\frac{\mathcal{G}_{12}}{\mathcal{G}_{21}}\right)^{\frac{1}{2}}\frac{1+\Sigma_1}{B(\theta)^2}\right]\sin^2\theta+ V_2\left[\frac{V_1}{V_2}\left(\frac{\mathcal{G}_{21}}{\mathcal{G}_{12}}\right)^{\frac{1}{2}}\frac{1+\Sigma_2}{B(\theta)^2}\right]\cos^2\theta - \sqrt{V_1V_2}\left[\frac{g_E/\sqrt{\omega_1\omega_2}}{B(\theta)^2}\right]\sin 2\theta. \\
\end{split}
\end{equation}
where
\begin{equation}
\begin{split}
A(\theta) &= \alpha_+\frac{\sqrt{V_1 V_2}}{V_+}=\sqrt{\textrm{Max} \left\{\varepsilon(\mathbf{x})\left[  \left(\frac{\mathcal{G}_{12}}{\mathcal{G}_{21}}\right)^{1/4}\sqrt{\frac{V_2}{V_1}}\,\mathbf{f}_1(\mathbf{x})\cos\theta+ \left(\frac{\mathcal{G}_{21}}{\mathcal{G}_{12}}\right)^{1/4}\sqrt{\frac{V_1}{V_2}}\,\mathbf{f}_2(\mathbf{x})\sin\theta\right]^2\right\}}\\
B(\theta) &= \alpha_-\frac{\sqrt{V_1 V_2}}{V_-}=\sqrt{\textrm{Max} \left\{\varepsilon(\mathbf{x})\left[\left(\frac{\mathcal{G}_{21}}{\mathcal{G}_{12}}\right)^{1/4}\sqrt{\frac{V_1}{V_2}}\,\mathbf{f}_2(\mathbf{x})\cos\theta  -\left(\frac{\mathcal{G}_{12}}{\mathcal{G}_{21}}\right)^{1/4}\sqrt{\frac{V_2}{V_1}}\,\mathbf{f}_1(\mathbf{x})\sin\theta\right]^2\right\}}
\end{split}
\end{equation}

\subsection{Effective Hamiltonian approach for calculation of transmission spectra}
\noindent Computation of the power transmitted through the coupled ring-nanobeam system is most easily achieved in the basis of creation and annihilation operators. Standard canonical quantization techniques rely on computation of the Hamiltonian associated with the Lagrangian in Eq. (\ref{eq:L}). Due to the coupling between $\dot{q}_1$ and $\dot{q}_2$, however, this leads to conjugate momenta which themselves are coupled in the Hamiltonian. The result of this is that the rotating-wave approximation is no longer valid, and standard techniques of input-output theory for coupled systems becomes ineffective.

The most straightforward path to quantization is via the effective Lagrangian
\begin{equation}
L = \frac{1}{2}\sum_i \left[\frac{\dot{q}_i^2}{\mathcal{V}_i} - \Omega_{i}^2\frac{q_i^2}{\mathcal{V}_i}\right] - \sqrt{\frac{\mathcal{G}_{12}\mathcal{G}_{21}}{\mathcal{V}_1\mathcal{V}_2}}q_1 q_2
\end{equation}
where $1/\mathcal{V}_1=\mathcal{G}_{21}/\sqrt{\bar{V}_1\bar{V}_2}$ and $1/\mathcal{V}_2=\mathcal{G}_{12}/\sqrt{\bar{V}_1\bar{V}_2}$. While different in form from the standard Lagrangian in Eq. (\ref{eq:L}), application of the Euler-Lagrange equations yields the exact same equations of motion. Notably, there is no direct coupling between $\dot{q}_1$ and $\dot{q}_2$, significantly simplifying quantization.

Legendre transform of the above Lagrangian yields the effective Hamiltonian
\begin{equation}
\begin{split}
H &= \sum_i\left[\frac{\mathcal{V}_i}{2}p_i^2 + \frac{\Omega_i^2}{2\mathcal{V}_i}x_i^2\right]+\sqrt{\frac{\mathcal{G}_{12}\mathcal{G}_{21}}{\mathcal{V}_1\mathcal{V}_2}}q_1 q_2 \\
&= \sum_i\hbar\Omega_i a_i^{\dagger}a_i + \hbar\sqrt{\frac{\mathcal{G}_{12}\mathcal{G}_{21}}{4\Omega_1\Omega_2}}(a_1^\dagger a_2 + a_1 a_2^\dagger)
\end{split}
\end{equation}
where $a_i=\sqrt{\Omega_i/2\hbar\mathcal{V}_i}[x_i + i(\mathcal{V}_i/\Omega_i)p_i]$ and counter-rotating terms have been discarded in accordance with the rotating wave approximation. This procedure also allows us to identify $\sqrt{\mathcal{G}_{12}\mathcal{G}_{21}/4\Omega_1\Omega_2}$ as the ``effective coupling strength'' to be compared with the dissipation rates in quantitatively testing for strong coupling. The transmission spectrum may then be computed through standard input-output methods \cite{haus1984waves, Collett1984}, yielding
\begin{equation}
P_{\textrm{tr}}(\omega) = \left|\frac{\kappa}{\omega-\Omega_1 + i\kappa+\frac{\mathcal{G}_{12}\mathcal{G}_{21}/4\Omega_1\Omega_2}{\omega-\Omega_2}}\right|^2.
\end{equation}

\bibliography{../../../JabRef/MasterBib}